\documentstyle[12pt,aaspp4]{article}
  
\slugcomment{ADP-AT-99-12, ~MPIfR No.~809, ~~Ap.J.(Lett.) accepted 1999}

\def\ltsima{$\;\buildrel < \over \sim \;$} \def\simlt{\lower.5ex
\hbox{\ltsima}} \def\gtsima{$\;\buildrel > \over \sim \;$}
\def\simgt{\lower.5ex \hbox{\gtsima}}
\begin{document}
  
\title{TeV Cherenkov Events as Bose-Einstein Gamma Condensations}
  
\author{Martin Harwit$^1$, R. J. Protheroe$^2$, and P. L. Biermann$^{3,4}$}
  
\affil{$^1$511 H Street S.W., Washington DC 20024--2725; also
Cornell University}
  
\affil{$^2$Department of Physics and Mathematical Physics,\\ The
University of Adelaide, SA 5005, Australia}
  
\affil{$^3$Max-Planck Institut f\"{u}r Radioastronomie, Auf dem
H\"{u}gel 69, D-53121 Bonn, Germany}
  
\affil{$^4$Department of Physics and Astronomy, University of
Bonn, Bonn, Germany}
  
\begin{abstract}  
The recent detection of gamma radiation from Mkn 501 at energies
as high as $\sim 25$~TeV suggests stringent upper bounds on the
diffuse, far infrared, extragalactic radiation density. The
production of electron-positron pairs through photon-photon
collisions would prevent gamma photons of substantially higher
energies from reaching us across distances of order
100\,Mpc. However, coherently arriving TeV or sub-TeV gammas ---
Bose-Einstein condensations of photons at these energies ---
could mimic the Cherenkov shower signatures of extremely
energetic gammas. To better understand such events, we describe
their observational traits and discuss how they might be
generated.  
\end{abstract}
  

\section{Introduction}
  
High energy gamma rays are readily absorbed in the intergalactic
medium through pair production in a sufficiently dense, diffuse,
microwave or infrared radiation field (Gould \& Schr\'eder,\
1966; Stecker, De Jager, \& Salamon\ 1992). For this reason, a
great deal of attention has be paid to gamma rays at energies
apparently reaching $\gtrsim 10$\,TeV, recently detected from the
galaxy Mkn 501 (Hayashida et al., 1998, Pian et al., 1998,
Aharonian et al., 1999, Krennrich, et al., 1999).  Mkn 501 is a
BL Lac object at a distance of $\sim 200$\,Mpc, for a Hubble
constant, H$_0$ = 50 km\,s$^{-1}$\,Mpc$^{-1}$.  Unattenuated
transmission of $\gtrsim 10$\,TeV photons across distances of
this order would place severe constraints on the diffuse
extragalactic infrared background radiation (Coppi \& Aharonian,\
1997, Stanev \& Franceschini,\ 1998) placing upper limits to the
radiation density that are close to values derived from COBE
detections and IRAS source counts alone (Hauser, et al., 1998;
Hacking \& Soifer, 1991; Gregorich, et al., 1995). Given these
close coincidences it is useful to re-examine the severity that
these observations place on the density of the diffuse
extragalactic infrared radiation (DEIR).

\section{Bose-Einstein Condensations of Photons}
  
Coherent radiation, i.e. highly excited quantum oscillators, are
produced in a variety of processes, but are also regular
components of blackbody radiation in the Rayleigh-Jeans tail of
the energy distribution. These excited oscillators correspond to
densely occupied radiation phase cells --- a Bose-Einstein
condensation of photons all having quantum-mechanically
indistinguishable properties, i.e. identical momenta, positions,
polarizations, and directions of propagation, within the
Heisenberg uncertainty constraints.
  
Given that cosmic ray particles can have energies going up to
$3\times 10^{20}$\,eV, and given that one expects a cutoff for
gammas from Mkn 501 at energies many orders of magnitude lower,
around 10 or 20\,TeV, it does not seem far-fetched to think that
the actually observed gammas reaching Earth might lie far out in
the low-frequency tail of some significantly more energetic
radiation field characterized by an equivalent temperature much
higher than a few TeV.
  
If this were the case, we would expect that the radiation
arriving at Earth could be highly coherent, meaning that phase
cells would be filled to rather high occupation numbers, $N$.  As
they interact with the DEIR, densely filled phase cells can
decline in population and lose energy only by going stepwise from
an initial occupation number $N$, to $(N-1)$, and from there to
$(N-2)$, etc. Because the mean free path for interactions of
photons with the DEIR is energy dependent, a fraction of a
coherent assembly of photons could penetrate appreciably greater
distances through the diffuse extragalactic radiation field than,
say, a single photon of the same total energy.
  
A number $N_a$ of such arriving photons, each with energy $h\nu$
would impinge on the Earth's atmosphere at precisely the same
instant, and would interact with the atmosphere producing an air
shower that emits Cherenkov light that could mimic that due to a
single photon with energy $N_ah\nu$ impinging on the atmosphere.
These two kinds of impacts could be distinguished by shower
images they produce and probably also by the fluctuations in the
energy distribution observed near the cut-off energy $E_{\rm CO}$
for a series of Cherenkov events.
  
Because of their high momenta, the arriving bunched photons would
spread over only the smallest distance $\Delta y$ in their
traversal through extragalactic space, given by the uncertainty
relation $\Delta p_y\Delta y \sim h$, where $\Delta p_y$ is the
uncertainty in transverse momentum. $\Delta p_y$ is the product
of the photon momentum $h\nu/c$ and the angular size that the
source subtends at Earth. The smallest dimension we could expect
would be of the order of an AGN black hole Schwarzschild radius
$\sim 3 \times 10^{13} M/(10^8 M_\odot)$~cm. This would make
$\Delta y \sim (10^8 M_\odot/M)10^{-3}$~cm --- negligible in
Cherenkov detection.

\section{Interpretation of Cherenkov Radiation Data}
  
TeV $\gamma$-rays are detected through the Cherenkov radiation
generated in the Earth's atmosphere by electrons in an ``air
shower'' initiated by the $\gamma$-ray. Such air showers are
electromagnetic cascades involving pair production and
bremsstrahlung interactions. As long as the energy of the photon
entering the atmosphere is sufficiently high, the Cherenkov yield
of the air shower is sensitive primarily to the total energy
deposited, not to the number of instantaneously arriving
photons. Accordingly, one might expect such telescopes to
mistakenly record five simultaneously arriving 5\,TeV photons as
a single shower of 25\,TeV. On the other hand, if the number of
simultaneously arriving photons, $N$, were much higher, then the
showers would look very different, and if $N$ were really large
there would be no Cherenkov radiation at all.
  
To quantify the discussion above, we shall compare the mean and
standard deviation of the number of electrons in the shower,
$N_e(t)$, as a function of depth into the atmosphere measured in
radiation lengths, $t$, for the two cases. Note that the
atmosphere is approximately 1030~g~cm$^{-2}$ thick and the
radiation length of air including forward scattering is
36.66~g~cm$^{-2}$. Although the cross section for interaction of
an assembly of $N$ coherent photons is $N$ times higher than that
of an individual photon, a shower initiated by an assembly of $N$
coherent photons having total energy $N \varepsilon$ would be
identical to a superposition of $N$ showers due to individual
photons of energy $\varepsilon$.  Above $\sim 3$~GeV the pair
production mean free path for photons in air is constant at
$t_{\rm pair}=9/7$ radiation lengths. For an assembly of $N$
coherent photons, the pair production mean free path is therefore
identical to an exponential distribution with mean $t_{\rm
pair}=(9/7)$, i.e. it is the same as the distribution of first
interaction points of a single photon. This also implies that at
depth $t$ the average number of photons remaining in the assembly
is $N \exp(-t/t_{\rm pair})$.
  
\begin{figure}
\plotone{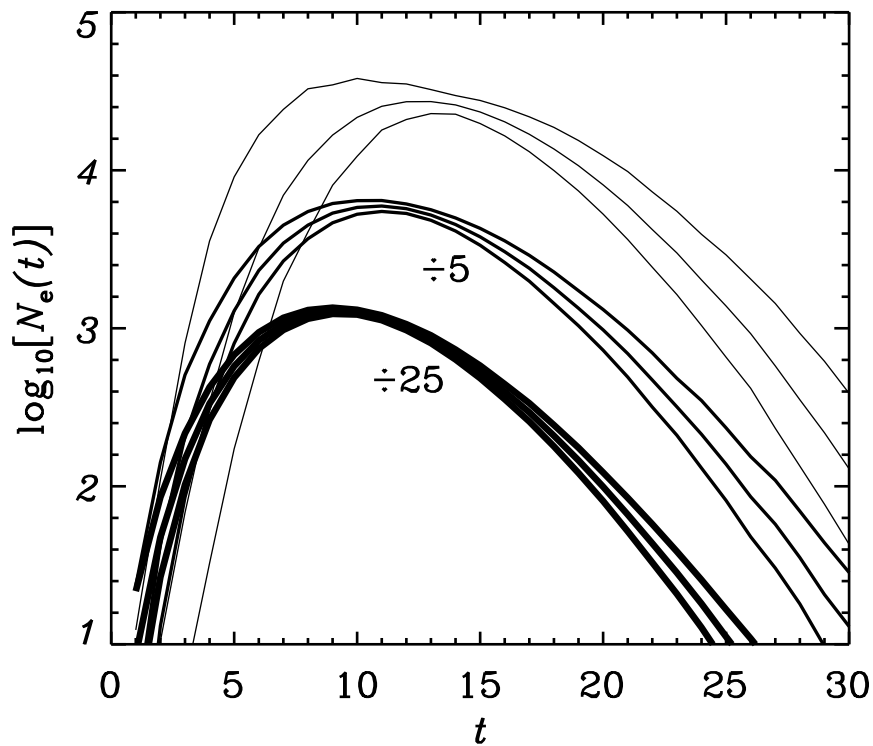}
\caption{Number of electrons, $N_e$, in the air shower versus
depth in the atmosphere measured in radiation lengths. The top
three curves show $\bar{N_e}$ and $\bar{N_e} \pm 1 \sigma$ for
single photons of energy 25~TeV, the middle three curves show
$\bar{N_e}$ and $\bar{N_e} \pm 1 \sigma$ for an assembly of five
coherent photons each of energy 5~TeV (note these curves have
been displaced down by a factor of 5 to avoid confusion), and the
bottom three curves correspond to an assembly of twenty-five
coherent photons each of energy 1~TeV (displaced down by a factor
of 25).
\label{fig1}}
\end{figure}

Crewther and Protheroe (1990) provide a parametrization of the
distribution of the number of electrons in photon initiated
showers, $p[N_e(t-t_1)]$, as a function of depth into the
atmosphere beyond the first interaction points of the primary
photons, $(t-t_1)$. We use their results together with our Monte
Carlo simulation of the first interaction points of each of the
$N$ photons in a coherent assembly to simulate the development of
the air shower due to the coherent assembly, thus taking account
of all fluctuations in shower development. In Fig.~\ref{fig1} we
show as a function of atmospheric depth $t$ $\bar{N_e}$ and
$\bar{N_e} \pm 1 \sigma$ based on 1000 simulations for the case
of single photons of energy 25~GeV, assemblies of 5 coherent
photons each having energy 5~TeV, and assemblies of 25 coherent
photons each having energy 1~TeV (each assembly has energy
25~GeV). As can be seen, air showers due to coherent assemblies
develop higher in the atmosphere, and have much smaller
fluctuations in shower development. Such differences between
showers due to single photons and assemblies of coherent photons
would produce different Cherenkov light signatures and should be
detectable with state-of-the-art Cherenkov telescopes such as
HEGRA (see e.g.  Konopelko et al. 1999).

\section{Extragalactic Optical Depth}
  
Propagation of assemblies of $N_0$ coherent photons each of
energy $\varepsilon$ through the microwave and DEIR fields is
analogous to their propagation through the atmosphere.  However,
assemblies of coherent photons having total energy $E_{\rm
tot}=N_0\varepsilon$ may travel farther than single photons of
energy $N_0\varepsilon$ without interaction because, unlike in
the atmosphere, the mean free path for pair-production in the
extragalactic radiation fields depends strongly on photon energy.
  
\begin{figure}
\plotone{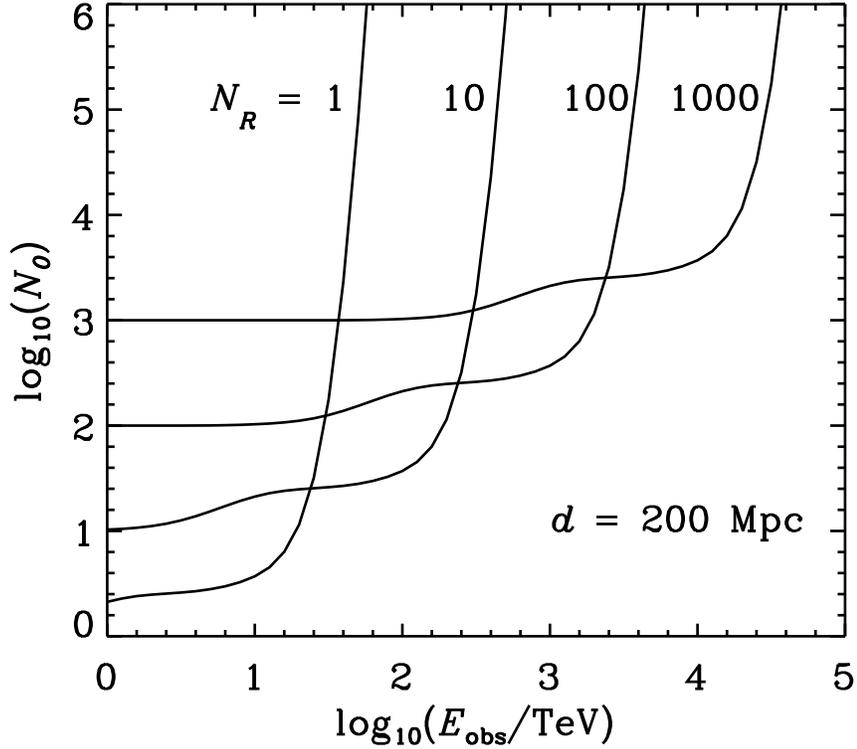}
\caption{Propagation of coherent photons across 200~Mpc through
the microwave and diffuse infrared extragalactic radiation
fields.  $N_0$ is the number of photons in the coherent assembly
emitted in our direction. $N_R$ is the surviving number. $E_{\rm
obs}$ is the energy of the observed Cherenkov event, equivalent
to $N_R \varepsilon$, where $\varepsilon$ is the energy of
individual photons in the assembly.
\label{fig2}}
\end{figure}

Just as in the air-shower cascade, only a single photon at a time
can be lost from a phase cell, with a corresponding decline in
occupation number from $N$ to $N-1$. On each encounter with an
infrared photon, the coherent assembly of $N$ photons has an
$N$-fold increase in probability for some photon to be removed,
so the mean free path is $x_{\rm pair}(\varepsilon)/N$ where
$x_{\rm pair}(\varepsilon)$ is the mean free path for
photon-photon pair production by single photons of energy
$\varepsilon$ through the extragalactic radiation fields. This
implies that at distance $x$ from the source the average number
of photons remaining in the assembly is $N_R(x)=N_0
\exp[-x/x_{\rm pair}(\varepsilon)]$, precisely the expression
that would hold for $N_0$ independent photons.  If $d$ is the
distance from the source to Earth, then the energy observable by
Cherenkov telescopes is $E_{\rm obs}=N_R(d) \varepsilon$, and the
number of photons in the assembly of coherent photons on emission
was $N_0 = N_R \exp[d/x_{\rm pair}(E_{\rm obs}/N_R)]$.  For the
purpose of illustration, we use for $x_{\rm pair}(E )$ the
logarithmic mean of the upper and lower curves of Fig.~1(a) of
Bednarek and Protheroe (1999) which is based on the infrared
background models of Malkan and Stecker (1998). We show in
Fig.~\ref{fig2} the result for propagation of coherent photons
through the microwave and DEIR fields across $d=200$~Mpc
appropriate to Mkn~501. We note, for example, that a coherent
assembly of forty 10~TeV photons emitted would typically arrive
at Earth as a coherent assembly of ten 10~TeV photons with an
observable energy of 100~TeV, while a single photon of 100~TeV
would have a probability of much less than $10^{-6}$ of reaching
Earth.

\section{Fluctuations in the Arriving Phase Cell Energy Content}
  
A stream of photons characterized by a brightness temperature
$T_b$ of necessity will also have a distribution of phase cell
occupation numbers, $N$, which, for high average values $\langle
N \rangle$ fluctuates as $(\Delta N)_{\rm rms} \sim \langle N
\rangle$. For emission of a stream of identical assemblies of
coherent photons, each containing $N_0$ photons on emission,
fluctuations in the number of photons, $N_R$, remaining in each
assembly after propagation to Earth through the DEIR, are
Poissonian about the mean value $\langle N_R \rangle$,
i.e. $(\Delta N)_{\rm rms} \approx \sqrt{\langle N_R \rangle}$,
for $\langle N_R \rangle \ll N_0$, and less than Poissonian for
$N_R \sim N_0$. Both these effects broaden the energy
distributions of observed Cherenkov events.

\section{What Mechanisms Could Produce Coherent TeV Gammas?}
  
In the laboratory (such as DESY), coherent X-radiation can be
produced by stimulated emission of relativistic electrons through
a periodically varying magnetic field (Madey, 1971).  Therefore
this shows that such processes are available in principle.
  
A more promising astrophysical process might arise from the
interaction of a collimated beam of relativistic electrons moving
roughly upstream against an OH or H$_2$O megamaser. This process
is attractive, because a substantial number of AGNs are known to
have nuclear megamasers. Inverse Compton scattering would produce
photons with an energy increase $\gamma^2$ in the co-moving frame
of the jet of relativistic, randomly directed electrons.  Here
$\gamma$ is the Lorentz factor of electrons in the jet's
co-moving frame. To produce 1\,TeV photons from H$_2$O megamaser
radiation at 22\,GHz, we would require $(\delta \gamma)^2 \sim
1.1 \times 10^{16}(E/{\rm TeV})$, where $\delta = [\Gamma( 1 -
\beta \cos \theta)]^{-1}$ is the Doppler factor, $\beta = v/c$
refers to the relativistic bulk velocity $v$, and $\Gamma \equiv
[1 - v^2/c^2]^{-1/2}$ is the Lorentz factor of the jet. The
factor $\delta^2$ translates the photon's initial energy to the
co-moving frame and back to the frame of an Earth-based
observer. For Mkn 501, the line of sight angle $\theta$ appears
to be directed very nearly in our direction, so we may choose
$\delta = 2\Gamma \sim 25$ (e.g.  Tavecchio et al. 1998).
  
As shown below, the number of phase cells into which the maser
photons can be inverse-Compton scattered is limited and quickly
fill up for relativistic jets with high column densities. At the
photon densities discussed, nonlinear effects can be neglected.
  
To provide a representative example, we might cite conditions in
the galaxy NGC 1052, which contains a water megamaser with
components that appear to lie along the direction of a radio jet
(Claussen et al. 1998). Though this may just be a projection
effect, we will assume as these authors have that it may signify
interaction of the jet with dense clumps of molecular clouds --
possibly producing maser activity in shocks.
  
The observed radiation intensity of the maser per unit bandwidth
at 22\, GHz is $I(\nu) = (c\rho(\nu)/4\pi) = 50 \; {\rm mJy}$ for
a beam size that is unresolved at $0.3\times 1$\,mas.  The beam,
however, is clearly much larger than the roughly forty individual
sources that are detected by virtue of their different velocities
along the line of sight, whose centroids are separated by as
little as $\sim 0.1$\,mas. The brightness temperature of these
individual sources is $T_b(\nu) \equiv [I(\nu) c^2/2 k\nu^2] >
4.5 \times 10^8 \; {\rm K}$ if the nominal beam size is
assumed. The density of phase space cells at this frequency is
$n(\nu) = (8\pi\nu^2/c^3) \sim 4.5\times 10^{-10} \; {\rm
cm}^{-3}\,{\rm Hz}^{-1}$ so that the phase cell occupation number
becomes $N_{\rm occ} = [\rho(\nu)/h\nu n(\nu)] = (kT_b/h\nu) >
4.3\times 10^8$.
  
All these figures are lower limits, since neither the angular
resolution nor the spectral resolution suffice to resolve the
individual maser sources. For this reason, it may be better to
assume the properties of the better-resolved Galactic H$_2$O
masers, which have a brightness temperature of order $T_b\sim
10^{14}\,$K, and a corresponding occupation number of order
$N_{\rm occ}\sim 10^{14}$ (Moran, 1997). To be somewhat more
conservative, we will adopt a value of $N_{\rm occ}\sim 3\times
10^{13}$ below.
  
Under a Lorentz transformation $I(\nu)$ and $\rho(\nu)$ scale as
$\nu^3$, as does $h\nu n(\nu)$, so that the phase cell occupation
number transforms as a constant. We can therefore deal with the
occupation number as though it were in the rest frame of the jet
of relativistic electrons. These electrons with energy $\gamma
m_ec^2$ will have some velocity dispersion, leading to an energy
bandwidth $\Delta\gamma m_ec^2$. On inverse-Compton scattering
the effective occupation number of scattered photons will be
reduced by the ratio of bandwidths,
$(\Delta\gamma/\gamma)/(\Delta\nu/\nu)$. If we take
$(\Delta\gamma/\gamma) \sim 1$, and $(\Delta\nu/\nu) \sim 3
\times 10^{-6}$ corresponding to a 1 km s$^{-1}$ velocity spread,
the reduction in occupation number is of order $3 \times 10^5$
bringing the actual occupation number down to $\sim 10^8$.
  
The occupation number of inverse-Compton scattered photons also
could in principle be diluted by the low, effective cross section
for back-scatter, i.e. by the Klein-Nishina cross section for
back-scattering. However, despite the $\gamma\delta$ value of
$10^8 (\delta/25)(\gamma/(4\times 10^6))$ for electrons, the
incident photons only have energy $\gamma h\nu\delta\sim
10^4$\,eV in the electron's rest frame, far lower than the
0.511\,MeV electron rest mass. The Klein-Nishina cross section,
therefore, reduces to the Thomson cross section $\sigma_T \sim
6.6 \times10^{-25}$\,cm$^2$.
  
We can assume that the masers are isotropic, or else, if they are
not, that there are a larger number than are actually
observed. Either way, the scattered light they produce would be
the same. If we further assume a jet with relativistic electron
column density through which the maser photons pass of order
$n_{e, rel} \ell \sim 10^{17}$\,cm$^{-2}$, we can estimate the
phase cell occupation number of the scattered radiation. It is
the product of the maser beam phase-cell occupation number, the
ratio of bandwidths, the electron column density, and the Thomson
cross section, giving
  
\begin{equation}  
N_{\rm occ}^{\rm scat} \sim 6 \biggl (\frac{N_{\rm occ}}{3\times
10^{13}}\biggr )\biggl
(\frac{(\Delta\nu/\nu)/(\Delta\gamma/\gamma)}{3\times
10^{-6}}\biggr ) \biggl ( \frac{n_{e, rel} \ell}{ 10^{17}{\rm
cm}^{-2}}\biggr )
\end{equation}
  
Interestingly, those phase cells with high back-scattered
occupation number $N_b$ will increase their occupancy at a rate
$(N_b + 1)$ times faster than unoccupied cells, since induced
scattering then begins to play a role --- there is gain. We may,
therefore, expect such a configuration to give rise to reasonably
high occupation numbers for TeV photons and energy densities
compatible with observed values.  NGC 1052, exhibits nearly 40
maser hot spots, with a total 22\,GHz luminosity of $\sim
200L_{\odot}\sim 8 \times 10^{35}$erg s$^{-1}$.  Let us assume
that the maser power available for interacting with the
relativistic jet would be equivalent to only 25\% of this. If
only fraction $n_{e, rel} \ell \sigma_T \sim 6.6 \times 10^{-8}$
of this radiation is scattered, but each photon's energy
increases by $\sim 1.1 \times 10^{16}$, the 1\,TeV luminosity is
$\sim 1.5 \times 10^{44}$\,erg\,s$^{-1}$. This needs to be
compared to the TeV flux from Mkn 501 in its high state, which is
of order $3 \times 10^{-10}$\,erg\,cm$^{-2}$\,s$^{-1}$,
corresponding for a distance of 200\,Mpc to an apparent
omnidirectional luminosity of $1.5 \times 10^{45}$\,erg\,s$^{-1}$
(Pian et al.  1998). Since our model assumes only a single jet
spherically expanding within a relatively narrow solid cone whose
axis is directed at us, these two figures are roughly consonant.

\section{Synchrotron Emission from Relativistic Electrons}
  
A highly relativistic electron with energy $E$ emits synchrotron
power in its rest frame $P(E) \sim 2.6 \times10^{-4} (B/0.1 {\rm
gauss})^2 (\gamma/4 \times 10^6)^2 {\rm erg\,s}^{-1}$.
  
The peak frequency the photons attain in this frame will be of the order of
  
\begin{equation}  
\nu_m \sim \frac{eB\gamma ^2}{2\pi m_ec} \sim 4.5\times 10^{18} \biggl
(\frac{B}{0.1 {\rm gauss}} \biggr )\biggl (\frac{\gamma}{4\times
10^6} \biggr)^2 {\rm Hz}
\end{equation}
  
\noindent where $m_e$ is the electron rest mass, $B \sim
0.1$~gauss (e.g. Bednarek \& Protheroe 1999) is the local
magnetic field strength, and $e$ the electron charge. In the
terrestrial observer's frame the frequency becomes
$\nu_m\delta\sim 10^{20}$\,Hz, which roughly corresponds to the
peak synchrotron radiation frequency of Mkn 501 in the high
state.
  
OSSE observations during flaring (Catanese et al. 1997) show that
the energy flux per log energy interval continues up to $\sim
500$ keV at roughly the same level as that observed by Beppo-SAX
(Pian et al. 1998), indicating that Mkn 501 emits a synchrotron
power at 0.5 MeV comparable to the TeV power during flaring. The
emitted synchrotron power in the relativistic jet's comoving
frame would be $2.4 \times 10^{42} (25/\delta)^2$\,erg\,s
$^{-1}$, implying emission from $\sim 10^{46}(25/\delta)^2(0.1
{\rm gauss}/B)^2(4\times 10^6/\gamma)^2$ relativistic electrons.
  
In recent models of AGN jet dynamics (e.g. Falcke \& Biermann
1999) a relativistic jet can readily interact with $N_{\rm cl}$
dense ambient molecular clumps located at $\sim 10^{19}\,$cm from
the central engine, to produce relativistic shocks that could
trigger maser emission in these clumps. Local acceleration at the
shock fronts or production from hadronic interaction and decays
could then also provide relativistic particle energies $\gamma
m_e c^2\sim 3.2(\gamma/4\times 10^6)$\,erg in the jet's comoving
system. The time scale for energy loss for these particles
through synchrotron radiation is of order $t_{synch}\sim
1.25\times 10^4 (4\times 10^6/\gamma)(0.1\,{\rm
gauss}/B)^2$\,seconds. Since the relativistic shocks propagate
into the jet at a significant fraction of the speed of light, the
radiating post-shock volumes have dimensions of order
$ct_{synch}\sim 10^{14}$ to $10^{15}$\,cm on a side. At particle
densities of order $n_{e, rel}\sim 10^2/N_{\rm cl}\,$cm$^{-3}$, a
post-shock column density of $10^{17}$ cm$^{-2}$, through the
$N_{\rm cl}$ shocks, therefore, appears possible.
  
\section{Discussion}
  
It is possible that highly energetic gamma radiation from distant
cosmological sources will be found to appear in conflict with
pair-production constraints imposed by the diffuse extragalactic
infrared background radiation. This apparent violation could then
be due to coherent TeV gammas of lower energy, whose Cherenkov
radiation superficially mimics individual photons of much higher
energy. We have suggested how the Cherenkov radiation signatures
of coherent and incoherent radiation can be distinguished, and
have sketched a plausible way in which coherent TeV photons could
be astrophysically generated. Whether this particular mechanism
is found in nature, remains to be determined, but other possible
sources of coherent TeV gamma radiation are also entirely
possible. If coherent TeV photons can be produced in nature then
we have shown that there exists a mechanism by which multi-TeV
Cherenkov signals may be observed from high redshift sources.
  
The work of one of us (MH) is supported by grants from NASA. The
Alexander von Humboldt Foundation, the Max Planck Institute for
Radio Astronomy in Bonn, and the Australia Telescope National
Facility were his gracious hosts during work on this paper. Drs.
Vladimir Strelnitski and Karl Menten kindly provided helpful
comments. The work of RJP is supported by the Australian Research
Council. PLB's work on high energy physics is partially supported
by a DESY grant. He wishes to acknowledge discussions with
Dr. Carsten Niebuhr of DESY, Hamburg, Dr. Yiping Wang of PMO,
Nanjing, and Dr. Heino Falcke and Ms.  Giovanna Pugliese from
Bonn.
  
\vfill \eject
  
\centerline{\bf References}
  
\vskip 0.1 true in
  
{\hoffset 20pt \parindent = -20pt
  
Aharonian, F. A., et al. for the HEGRA collaboration 1999 A \& A submitted,
astro-ph/9903386
  
Bednarek, W. \& Protheroe, R. J. 1999, MNRAS, in press astro-ph/9902050
  
Catanese M. et al., 1997 ApJ, 487, L143
  
Claussen, M. J., et al. 1998, ApJ, 500, L129
  
Coppi, P. S. \& Aharonian, F. A. 1997, ApJ, 487, L9
  
Crewther, I. Y. \& Protheroe, R. J. 1990, J. Phys. G:
Nucl. Part. Phys., 16, L13
  
Falcke, H. \& Biermann, P.L. 1999, A\&A, 342, 49
  
Gould, R. J. \& Schr\'eder, G. 1966, PRL, 16, 252
  
Gregorich, D. T., et al. 1995, AJ, 110, 259
  
Hacking, P. B. \& Soifer, B. T. 1991, ApJ, 367, L49
  
Hauser, M. G., et al. 1998, ApJ, 508, 25, astro-ph/9806167
  
Hayashida, N., et al. 1998, ApJ, 504, L71
  
Konopelko, A., et al. for the HEGRA collaboration 1999,
Astropart.  Phys., 10, 275
  
Krennrich, F., et al. 1999, ApJ, 511, 149
  
Madey, J. M. J. 1971 J. Appl. Phys., 42, 1906
  
Malkan M.A. \& Stecker F.W. 1998, ApJ, 496, 13
  
Moran, J. M. 1997, ``Modern Radio Science", ed. J. H. Hamelin,
International Radio Science Union (URSI), Oxford University Press
(also Harvard-Smithsonian Center for Astrophysics Preprint
No. 4305)
  
Pian, E., et al. 1998, ApJ, 492, L17
  
Stanev, T. \& Franceschini, A. 1998, ApJ, 494, L159
  
Stecker, F. W., De Jager, O. C., \& Salamon, M. H. 1992, ApJ, 390, L49
  
Tavecchio, F., Maraschi, L., \& Ghisellini, G. 1998, ApJ, 509, 608 }
  

\clearpage

\clearpage

\end{document}